\documentclass[preprint]{aastex}
\usepackage{graphics}
\usepackage{psfig}

\def\thread{{G359.54+0.18}}
\def\xthread{{G359.54+0.18}}

\def\g359{{G359.89-0.08}}
\def\G359{{G359.89-0.08}}

\begin{document}

\slugcomment{draft version}
\shorttitle{G359.89-0.08 \& G359.55+0.17}
\shortauthors{Lu et al.}

\title{The Chandra Detection of Galactic Center X-ray Features G359.89-0.08 and G359.54+0.18}


\author{F.J. LU\altaffilmark{1,2}, Q.D. WANG\altaffilmark{1}, and C.C. LANG\altaffilmark{3}}


\altaffiltext{1}{Astronomy Department, University of Massachusetts,
    Amherst, MA 01003; lufj@flamingo.astro.umass.edu; wqd@astro.umass.edu}
\altaffiltext{2}{Laboratory of Particle Astrophysics, 
Institute of High Energy Physics, CAS, Beijing 100039, P.R. China}
\altaffiltext{3}{Department of Physics \& Astronomy, University of Iowa, Iowa
City, IA 52245; cornelia-lang@uiowa.edu}

\begin{abstract}
We report on the detection of two elongated X-ray features  
\g359\ and \thread\ in the Galactic center (GC) region using the 
{\sl Chandra X-ray Observatory}.  \G359 is an elongated X-ray feature
located $\sim$2\arcmin~in projection south of the center of the Galaxy, SgrA$^*$.
This X-ray feature source is partially coincident with a slightly curved (``wisp''-like)
non-thermal radio source. The X-ray spectrum of 
\g359 can be best characterized as non-thermal, with a photon index of 2.  
The morphological and spectral characteristics of the X-ray and radio emission associated
with \g359 are best
interpreted as the synchrotron emission from a ram-pressure confined pulsar wind nebula. 
G359.54+0.18 is one of the most prominent radio non-thermal filaments (NTFs) in
the GC region, located $\sim$30\arcmin~in projection from SgrA$^*$. A narrow 
($\sim$10\arcsec~) filament of X-ray emission appears to arise from one of 
the two strands that comprise the radio NTF. Although the photon statistics 
are poor for this source, the X-ray emission is also likely to be non-thermal in nature.
Several models for the production of X-ray emission in \xthread\ are discussed.

\end{abstract}


\section{Introduction}

The central region of our Galaxy hosts many unique and filamentary radio 
features (e.g., Yusef-Zadeh, Morris, \& Chance 1984; Morris \& Serabyn 1996; 
Ekers et al. 1983; Anantharamaiah et al. 1991). Most prominent of these are the
non-thermal filaments (NTFs), the brightest of which extend for 
up to 20\arcmin~($\sim$50 pc at the 8.0 kpc distance of the Galactic center (GC)).  
Radio polarization measurements show that the NTFs represent 
synchrotron radiation from relativistic particles on local magnetic 
fields (Yusef-Zadeh \& Morris 1987; Lang, Morris \& Echevarria 1999). 

These NTFs are oriented in directions essentially 
perpendicular (within $\sim$20$\degr$) to the Galactic plane
and are thought to trace a large-scale poloidal magnetic field configuration in
the GC region. However, the origin of the magnetic field arrangement in the GC, 
and the source of particles and their acceleration in the NTFs 
remain unclear (e.g., Morris \& Serabyn 1996). Less extensive
versions of these NTFs are also apparent throughout the GC region, and 
show the same physical characteristics, but on smaller 
scales (Ho et al. 1985; Morris \& Yusef-Zadeh 1989; Lang, Morris \& Echevarria 1999; 
LaRosa et al. 2000; LaRosa, Lazio \& Kassim 2001). 

We have detected three elongated X-ray features in the 
recent {\it Chandra X-ray Observatory} GC survey 
(GCS), which covers the central $\sim$300 pc along the Galactic 
plane (Wang, Gotthelf \& Lang 2002a). Most 
interestingly, we find that the elongated X-ray features 
are apparently associated with either NTFs or NTF-like radio sources.
These apparent associations, if confirmed, have strong implications for
understanding the nature of the radio NTFs. 
In particular, if these X-ray features are also due to synchrotron 
emission, they directly trace the sites where relativistic particles
are freshly accelerated (because the X-ray particles have an extremely short lifetime).
Identifying such features would allow us to locate the origin
of the particle acceleration in the mysterious radio NTFs.

The analysis of one of these X-ray features, G0.13-0.11 has already been
presented (Wang, Lu, \& Lang 2002b). A detailed study of the morphology 
and spectrum of the X-ray
and radio emission suggest that G0.13-0.11 represents the leading-edge of a pulsar wind 
nebula (PWN), produced by a pulsar moving in a strong magnetic field
environment. The main body of this PWN is traced by a radio
protrusion that may also be related to the well-known nonthermal Radio Arc 
NTF (Yusef-Zadeh et al. 1984). Here, we present the analysis and interpretation of
two similarly-shaped X-ray features, \g359 and G359.54+0.18.
It should be noted that recently, Sakano et al. (2002)
have presented {\sl Chandra} and {\sl XMM-Newton} data on \g359. This source
has also been recorded by Baganoff et al. (2003) in their paper on the
{\it Chandra} observation of the central 15\arcmin~(38 pc) of the Galaxy. 
However, neither of the two papers has provided a complete 
physical interpretation of the detected X-ray emission in this unusual elongated X-ray feature.
In this paper, we present the detection ($\S$2), analysis ($\S$3) 
and discussion ($\S$4$-$$\S$6) of the X-ray emission in the features \g359 and G359.54+0.18.

\section{X-ray Observation and Data Reduction}

The X-ray data of \g359\ were obtained using 
the {\it Advanced CCD Imaging Spectrometer} (ACIS-I) onboard the {\it Chandra X-ray Observatory} 
in two observations pointed at Sgr A$^{*}$: OBS\#242 was taken on 
1999 September 21 with a total exposure time of 51.1 ks; OBS\#1561 consists
of a series of short exposures from 2000 October 26 to 2001 September 28 
with a total integration time of 49.8 ks. \g359\ 
is on the ACIS-I2 chip, with an off-axis angle of 3$\farcm$7.
\thread\ was covered by two of the {\it Chandra} GC Survey observations (OBS\#2268, and \#2275) (Wang et al. 2002a). 
OBS\#2268 observation was made on 2001 July 20 with a total integration 
time of 11.0 ks, and OBS\#2275 was made on 2001 July 20, with a total integration time of 12.0 ks. 
The diffuse X-ray emission feature is located on the ACIS-I0 chip in OBS\#2268, with an offset of 3$\farcm$5 from the aimpoint, and in OBS\#2273, the feature is located on the ACIS-I3 chip, with an offset of 7\arcmin~from the aimpoint. 

We calibrated the data using the most recent version of the {\it Chandra} data processing 
software {\it CIAO}. After excluding time intervals of high background we obtained 40.2 ks 
net exposure for OBS \#242, 35.7 ks for OBS \#1561, 10.9 ks for OBS \# 2268, and 
11.6 ks for OBS\#2275. While the pointing astrometry of OBS \#242 
is better than 1$\arcsec$ (Baganoff et al. 2003) and thus no
additional calibration is required, we corrected 
offsets, 0\farcs64 and 2\farcs05 in the R.A.  
and Dec, for OBS \#1561, and -0\farcs11 and 2\farcs12 in the
R.A. and Dec, for OBS \# 2268 and \#2275, using the online web 
tool ({\it http://asc.harvard.edu/cal/ASPECT/fix\_offset/fix\_offset.cgi}). 
The data from OBS \#242 and \#1561 were co-added in the sky coordinates 
to improve the counting statistics. 

The {\it Chandra} data are sensitive to the photons in the energy range
0.5-10 keV for OBS \#242 and \#1561 and 1-10 keV for OBS \# 2268 and \#2275 because
of an imposed lower energy telemetry cutoff (Wang et al. 2002a). The 
data, corrected for the CTI effect of the CCD chips with the Penn State
software (Townsley et al. 2000), provide a spectral resolution of $E/\delta E 
\sim 30$. The spatial resolution, a function of the location in the ACIS-I
CCD, is in the range of a couple of arcseconds for the X-ray features 
discussed here. Both the spatial and spectral resolution in this study 
are largely limited by the limited counting statistics of the data. 

\section{X-ray Properties}
\subsection{\G359} 

Figure 1 shows the count distribution in this field for energies of 4.0$-$9.0 keV. 
An elongated concentration of X-ray emission is present 
near R.A., DEC. (J2000) = 17$^{\rm h}$45$^{\rm m}$40$\fs$4, 
-29$\degr$04$\arcmin$29$\arcsec$, or {\it l,b} = 359.89, -0.08. Thus,
this source will be referred to as \g359. Figure 2 shows the 0.5-4.0 keV band count map overlaid with 
the contours of the smoothed count distribution in the 4.0-9.0 keV band. 
The elongation of \g359 occurs in the NW--SE direction with a length of 
$\sim$24$\arcsec$ and a width of $\sim8\arcsec$.   
We also detect a point-like source (CXOU J174539.6-290413; Figures 1 \& 2) 
with a total of 11 counts detected within an 1\farcs5 radius;  the 
expected number of the local 
background counts is $\sim 3$ and therefore this detection has 
a signal-to-noise ratio of $\sim 2.5$. A comparison of the count
rates of the source in the two observations (OBS \#242 and \#1561) shows no significant variability.

The energy distributions of \g359\ and CXOU J174539.6-290413 are similar, with 
counts almost entirely in the energy range above 4.0 keV as Figures 1 and 2
illustrate well. There is no significant soft X-ray enhancement ($<$4 keV) associated 
with either \g359\ or the point-like source. This implies that they are
at about the same distance and that the line-of-sight X-ray absorbing gas
column density may be quite large ($N_H > 10^{23}$ cm$^{-2}$).

We carried out a spectral analysis of \g359 by extracting the spectrum of the
extended source from the rectangular region illustrated in Figure 1. The
large box in Figure 1 represents the region used for determination of the
local background. Because of the limited counting statistics, the spectral fits
alone cannot be used to distinguish between models. For example, 
the spectrum of \g359\ can be fitted by either a thermal plasma (Raymond \& Smith 1977) or
a simple power law model, both with a strong interstellar medium (ISM) absorption.
Figure 3 shows the best-fit power law model to the data of \g359. The parameters of the
two different model fits are listed in Table 1.  
The best fit to the 
absorbing gas column density is 3.7$\times10^{23}$ cm$^{-2}$, which, for the power law model 
gives an X-ray (0.2-10 keV) luminosity of L$_x$ $\sim$
1.6$\times$10$^{34}$ $d_8^2$ erg s$^{-1}$, where $d_8$ is the distance in units of
 the distance of the GC, 8 kpc. 
There is some excess near 6.7 keV (Figure 3), however it is most likely due to
background noise fluctuations as this excess is not seen in the XMM data of \g359\
(Sakano et al. 2002). 
 
With only $\sim$11 total counts from the point-like source, CXOU J174539.6-290413, 
little can be said about its spectral shape. We thus assume a power law 
spectrum with a photon index of 2 and the same absorbing gas column density 
as the best-fit value obtained for \g359 (3.7$\times10^{23}$ cm$^{-2}$). 
The detected count flux of 1.4$\times10^{-4}${\rm~counts~s$^{-1}$} then gives 
a 0.2-10 keV X-ray luminosity of 8 $\times$10$^{32}$ $d_8^2$ erg s$^{-1}$, where $d_8$
is the distance in units of 8 kpc.

\subsection{\thread}

Figure 4 shows the X-ray emission that appears to be associated with 
the well-known radio NTF G359.54+0.18. This X-ray feature has a strikingly
linear morphology extending for nearly 2\arcmin, and corresponds exactly
with the brightest portion of the northern of the two radio filaments
in G359.54+0.18. While the width of the X-ray feature is broaden in the
smoothed image because of the limited counting statistics,
the boundary of the X-ray emission appears to 
correspond well to the width of the radio filament. This clearly suggests
that these two features are likely to be physically related. Figure 5 
shows the average (but unsmoothed) 
X-ray intensity as a function of distance perpendicular 
to its long axis (a cross-cut) from the raw counts image. The width of this 
X-ray filament is resolved to be $\sim 10$\arcsec. The distance of the
radio NTF G359.54+0.18 has now been constrained to be within a few hundred
parsecs of the GC (Roy 2003); therefore, the linear size of the X-ray emitting region is 
approximately 10 pc $\times$ 0.5 pc.  

The spectral shape of \thread~is more difficult to characterize
with the only $\sim$70 counts collected in the 
1-10 keV range.
We find that the spectrum of \thread~is consistent with a power law with a photon 
index equal to 2 and an X-ray-absorbing gas column density 
of $n_H\sim10^{23}$ cm$^{-2}$, typical for a source in the GC region.
The inferred X-ray luminosity of the filament, L$_x$(0.2-10 keV) is 
2.4$\times10^{33}$ $d_8^2$ erg s$^{-1}$.
Alternatively,  the spectrum can also be characterized by 
a Raymond-Smith (1977) thermal plasma of a temperature $\sim$ 2 keV and 
a similar luminosity as the above power law fit.

\section{Radio Properties}
Both \g359 and \thread~have prominent and unusual radio counterparts. We discuss the radio properties
of these sources in the following.

\subsection{\g359}

Previous radio data on \g359\ were originally obtained by 
Yusef-Zadeh \& Morris (1987) with the Very Large Array (VLA) radio telescope at 20 cm. These observations of 
this region, in addition to other VLA archive data, were reprocessed and imaged by Lang et 
al. (1999). For this study we have re-imaged both the Yusef-Zadeh \& Morris 
(1987) and Lang et al. (1999) data to make new estimates of the flux of 
radio features for comparison with the X-ray (see below).

Figure 6 shows an overlay of the X-ray emission in \g359 and the radio
continuum emission from the corresponding ``radio wisp'', first
recognized by  Ho et al. (1985) at 2 and 6 cm and was then shown 
at 20 cm (e.g., Yusef-Zadeh \& Morris 1987). Ho et al. (1985) suggested
that the wisp was part of a supernova remnant (SNR) shell with a 
nonthermal spectrum ($\alpha$$\sim$$-$0.4; where S$_{\nu}\propto\nu^{\alpha}$). 
However, this radio spectral index was not well determined (Ho et al. 1985). 
We have re-calculated the radio spectral index, using the radio data at 2 and
20 cm. We measured the integrated 20 cm flux of the wisp to be 720 mJy.
This flux, together with the 2 cm flux of the wisp of $\sim$447 mJy 
(Ho et al. 1985), suggests that the index is $\alpha$$\sim$$-$0.2.  
The radio luminosity ($10^7-10^{11}$ Hz, derived from the 2 cm flux density 
and using this spectral index of $-$0.2) is 2.9$\times10^{33}$ $d_8$ erg s$^{-1}$.

Interestingly, only part of the X-ray wisp \g359\ and the radio wisp overlap
spatially. The X-ray intensity centroid of \g359\ shows an
offset of $\sim$10$\arcsec$ toward the NW from the centroid of 
the radio wisp. Although this position offset needs to be explained, 
the radio wisp is  likely to be physically associated with \g359. 
In the 8$^{\prime}\times$6$^{\prime}$ region around the GC,
excluding the SgrA complex, only two isolated nonthermal diffuse 
radio features have been detected (Ho et al. 1985); the chance 
for a random superposition of the centroid of the X-ray wisp 
within the boundaries of one of these two nonthermal radio 
features is $\sim 5 \times10^{-3}$. Furthermore, the other two 
elongated X-ray features detected so far in the GCS (G0.13-0.11 (Wang et al. 2002b); G359.54+0.18, this paper) 
also have radio counterparts,
and we thus expect that the spatial and morphological coincidence between  \g359\ and 
the radio wisp represents a physical association.

Coil \& Ho (1999; 2000) have also argued for a possible association between
the radio wisp and a molecular ``streamer'', which is projected between 
the so-called 20 km s$^{-1}$ cloud M-0.13-0.08 and Sgr A$^*$. 
The lower tip of the streamer coincides spatially with the radio wisp and 
\g359 (see Figure 4 of Coil \& Ho (2000) and Figure 6 in this paper). 
The molecular hydrogen column density of the main body of
 this streamer is about 1.4$\times$10$^{24}$ cm$^{-2}$. Since 
the NH$_3$ emission strength at the lower tip is $\sim$1/5 of the NH$_3$ emission strength of
 the main body of the streamer (Coil \& Ho 1999; 2000), 
the molecular hydrogen column density along the line of sight to \g359\ is  
about 3$\times$10$^{23}$ cm$^{-2}$, consistent with the 
difference ($\sim$2.7$\times$10$^{23}$ cm$^{-2}$) between the X-ray absorbing 
column densities to \g359\ (3.7$\times10^{23}$ cm$^{-2}$) 
and to Sgr A$^*$ ($\sim$10$^{23}$ cm$^{-2}$) 
(e.g., Baganoff et al. 2001a).  Therefore, \g359\ is most likely 
located behind this streamer. We assume here that \g359\ has a 
distance comparable with that to the GC and scale the luminosities with $d_8$, 
the distance in units of 8 kpc (distance to GC).  

\subsection{\thread}
 
In the case of \thread, VLA observations of this source were originally made by Yusef-Zadeh, 
Wardle, \& Parastaran (1997) with the VLA at 3.6 and 6 cm. Their observations
(Figure 4) show that this radio NTF extends for 8\arcmin~(20 pc) and has a very 
prominent bi-furcated morphology, which is common of most of the GC NTFs.
The 90 to 20 cm spectral index of \thread~has been calculated to be 
$\alpha$=$-$0.8 (Anantharamaiah et al. 1991). The radio luminosity ($10^7-10^{11}$ Hz, 
(derived from a 6 cm flux density of 150 mJy from the region where the X-ray emission 
is arising and using $\alpha$=$-$0.8) is 5$\times10^{32}$ $d_8$ erg s$^{-1}$.
Similar to the other well-studied NTFs (e.g. Lang et al. 1999), G359.54+0.18 is characterized 
by strongly linearly polarized emission and a well-ordered magnetic 
field along its long axis (Yusef-Zadeh et al. 1997). 
The 6 to 3.6 cm spectral index is slightly steeper (by $\alpha$=$-$0.5) 
along the eastern side of the NTF than that of the center of the NTF and western
side, which may indicate a steepening in the energy distribution to the East.

\section{Nature of the X-ray Emission}

\subsection{Thermal Emission}
The spatial and spectral properties of the X-ray and radio emission
presented above provide useful constraints on the physical nature of
the X-ray emission in \g359 and \thread. In the case of \g359, the thermal hypothesis requires a  
temperature $\ge$13.5 keV (Table 1). No diffuse X-ray feature with
such a high thermal temperature is known in the Galaxy. If the X-ray emission is
really thermal, the closest 
possibility is that \g359\ could be a background cluster of galaxies, although
the temperature is still considerably higher than the typical cluster temperature
(2$\times10^7$--10$^8$ K; Fabian 1994). The small sizes and
high fluxes of the observed X-ray emission (Table 1) would further imply a
very distant rich cluster. Morphologically, such a cluster should appear 
relatively round, inconsistent with the wisp-like appearance of \g359.
This distant cluster hypothesis would also be difficult to interpret the 
high flux and offset morphology of the apparently associated radio emission.
We thus conclude that \g359\ is not a thermal source.

The thermal interpretation of the X-ray emission associated with G359.54+0.18 has the 
similar difficulties. In particular, the X-ray morphology of G359.54+0.18 is even 
more narrowly distributed than \g359. If the filament represents a cylinder with a diameter
of $\sim$10$\arcsec$,  the volume of the X-ray-emitting region is then 
$\sim$ 5$\times10^{55}$ cm$^{3}$ at the distance of 8 kpc. 
>From the X-ray luminosity, we can further infer the 
thermal pressure to be $\sim 8 \times10^{-9}$ dynes cm$^{-2}$. 
Because the magnetic field in the GC region is likely to be in the range of 
10$^{-4}$ to 10$^{-3}$ Gauss (e.g., Ananthanamaiah et al. 1991; 
Morris \& Serabyn 1996), the corresponding magnetic 
field pressure (4$\times10^{-10}$ to 4$\times10^{-8}$ dynes cm$^{-3/2}$) 
is probably not high enough to confine the thermal plasma in such a 
narrow region. Furthermore, the thermal origin of the X-ray-emitting
gas provides no explanation for the nonthermal radio emission, which is
much more extended linearly.
Therefore, a thermal origin of \thread~is problematic. 

\subsection{Nonthermal Emission}

Nonthermal X-ray emission in these two sources may arise from synchrotron 
emission or from inverse Compton scattering. The presence of the nonthermal
radio emission is a testimony of the presence of relativistic electrons.
Furthermore, it is important to note that the luminosity is considerably 
greater in X-ray than the radio for both G359.89-0.08 and G359.54+0.18 (\S  3-4).
 
The intensity of the inverse Compton scattering is proportional to the energy density of a 
photon field $U_{ph}$. In the GC, the energy density is generally 
dominated by Far-IR photons, $\sim 10^{-10} {\rm~ergs~s^{-1}}$ (e.g., Odenwald \& Fazio 1984), 
compared with the energy density of the microwave background 
radiation, $U_{ph} \sim 4 \times 10^{-13}$ erg cm$^{-3}$, or the radio 
synchrotron radiation contained in the radio NTFs and NTF-like features, 
$\sim$10$^{-16}$ erg cm$^{-3}$ (e.g. Yusef-Zadeh et al. 1997; Ho et al. 1985). 
The Far-IR energy density may be even comparable to 
the magnetic field energy density ($\sim$4$\times$10$^{-10}B_{-4}^2$ 
erg cm$^{-3}$, where $B_{-4}$ is the 
magnetic field strength in units of 0.1 mG; Morris and Serabyn 1996).
 
The intensity ratio of inverse-Compton scattering over synchrotron 
radiation is approximately $\sim {\gamma_c^{2-p} U_{ph} \over 
\gamma_s^{2-p} U_B} $, where $p$ is the index of the assumed power law 
particle energy distribution. The electron that scatters
a far-IR photon ($\sim 200$ $\mu$m) to X-ray ($\sim 4$ keV) would
typically have a Lorentz factor $\gamma_c \sim 10^2$, and is about an order 
of magnitude less energetic than the particles that are responsible for the 
observed radio synchrotron emission from the NTFs ($\gamma_s \sim 10^3$).
Therefore, in principle, the inverse-Compton scattering can produce
an X-ray intensity that rivals the synchrotron radiation, if $p = 2\alpha+1$ is
significantly less 2 and/or there is a local enhancement of the far-IR photon
field. But it is difficult for the inverse-Compton scattering mechanism 
to explain the fact that the X-ray arises only from part of the radio-emitting
regions, especially in the case of G359.54+0.18. The two strands in this NTF are
too close positionally for only one of them be affected by a local enhancement of the far-IR photon
field, for example. Although IC scattering can not be ruled out completely, it appears that synchrotron radiation 
provides a more natural interpretation of the X-ray emission of both \g359\ 
and \thread, which is detailed in the following.

\section{Origins of Relativistic Particles}

\subsection{\g359}
\subsubsection{Supernova Remnant (SNR) Origin}

The first possibility for the origin of synchrotron particles in \g359 is 
the nearby SNR (G359.92-0.09) (Ho et al. 1985; Coil \& Ho 2000). 
Evidence for this SNR is present in the morphology of the radio
features in this region, including the apparent circular structure consisting of 
the radio ``wisp'' and other shell-like radio sources (Ho et al. 1985). 
It has been proposed that the SNR shock could generate particles that may be energetic 
($\sim 10^{14}$ eV) enough to produce X-ray synchrotron radiation
(e.g., Koyama et al. 1995).

However, it is difficult to associate \g359\ with this SNR G359.92-0.09. 
Firstly, if the X-ray synchrotron emission arises from a SNR, a tight 
spatial correlation between the radio and X-ray emission would be expected if they both arise from
particles accelerated by the same SNR shock. Indeed,
such correlations are observed in all SNR shells that are dominated 
by nonthermal radio and X-ray emission, such as 
SN 1006 (Reynolds \& Gilmore 1986; Koyama et al. 1995; Willingale 1996)
and RCW 86 (Borkowski et al. 2001). The 10-30\arcsec~offset between
the radio and X-ray emission in \g359, is therefore unusual and difficult
to explain, although we can not exclude the possibility that the 
radio and X-ray features are from different parts of the SNR shell
and in projection, overlap. 

Secondly, the molecular line (NH$_3$) study of the central 15 pc region of the Galaxy also shows evidence
for the interaction between an SNR and the surrounding molecular streamers (Coil \& Ho 2000). 
The velocity gradients and velocity displacement of the southern tip of a 
molecular streamer thought to be interacting with \g359~suggests that the
SNR is partially in front of the molecular gas and likely to be embedded in a 
distribution of disturbed molecular material (Coil \& Ho 2000). 
However, the high X-ray-absorbing gas column density along the line of sight 
to \g359\ suggests that the X-ray emission must lie behind some enhancement of molecular
material ($\S$4) and is not consistent with an X-ray detection of a SNR shell or arc
that lies {\it in front} of the molecular gas. The X-ray emission in \g359\ 
is unlikely to be a detection of the SNR G359.92-0.09.

For these reasons, we conclude that  \g359\ is unlikely part of the SNR shell and
that the X-ray synchrotron emission does not arise from shocked-SNR particles.

\subsubsection{Pulsar Wind Nebula (PWN)}
We find that the X-ray properties of \g359\ are consistent with
what is expected for a PWN. Both the spectrum and the luminosity of \g359\ 
are quite typical for a PWN, which usually has a nonthermal X-ray spectrum with 
photon index 1.1 to 2.4 and 0.2-10.0 keV X-ray luminosities 
from 4$\times$10$^{32}$ to 2$\times$10$^{37}$ erg s$^{-1}$ 
(e.g., Gotthelf \& Olbert 2001). 

Naturally, we speculate that X-ray source CXCGCS J174539.6-290413 is the putative pulsar that powers the 
PWN. Indeed, the X-ray characteristics of this source, though weakly 
constrained, are consistent with being an isolated young pulsar at
the similar distance as \g359 (e.g., Becker \& Tr\"umper 1997). 
>From the X-ray luminosities of both the source and \g359, we estimate the 
spin-down power $\dot{E}$ of the pulsar as 
$\sim$4$\times10^{36}$ erg/s (Seward-Wang's 1988; Becker \& Tr\"umper 1997)

The wisp morphology of \g359\ and the location of the source 
CXCGCS J174539.6-290413 further suggest that they form a ram-pressure 
confined PWN  (e.g., Wang, Li, \& Begelman 1993).  
When a pulsar moves supersonically in a medium, a bow shock is 
expected to develop ahead of the pulsar. The pulsar wind
is thermalized at the reverse shock, which can lead to acceleration
of particles just as in a normal PWN. The ram-pressure of the bow shock
then sweeps up the shocked pulsar wind material into a trail in the
direction opposite to the pulsar motion. The pulsar wind material in 
this trail is relativistic plasma and can have a bulk motion that is mildly
relativistic. Such systems have been observed recently in N157B 
(Wang et al. 2001), SNR G5.3-1.0 (Kaspi et al. 2001), and SNR W44 
(Frail et al. 1996; Petre et al. 2002). 

>From the length of the X-ray feature $\sim$3 lyr sin$^{-1}\theta$ $d_8$, 
where $\theta$ is the angle between the flow direction and the line of sight,
we estimate the lifetime of the X-ray emitting 
particles ($\tau\sim40$ yr $B_{-4}^{-1.5}\epsilon^{-0.5}$, 
where $B_{-4}$ is the magnetic field strength in 
units of 10$^{-4}$ Gauss and $\epsilon$ is the photon energy in keV).
The bulk motion velocity is thus
about 0.15 $c$ $B_{-4}^{1.5}$sin$^{-1}\theta$$d_8$, in which
a photon energy of 4 keV has been used. With the magnetic field in 
the GC region in the range of 10$^{-4}$ to 10$^{-3}$ Gauss 
(e.g., Anantharamaiah et al. 1991, Morris \& Serabyn 1996),
the bulk velocity could be a significant fraction 
of the speed of light. 

The above ram-pressure confined PWN interpretation also provides
a natural explanation of the radio wisp and its offset from the
X-ray wisp \g359. The radio synchrotron particles have a much longer 
lifetime ($\tau\sim6\times10^5$ yr $\nu_{GHz}^{-0.5}B_{-4}^{-1.5}$),
comparable to, or less than, the age of the pulsar. Thus we expect
the radio synchrotron emission from the accumulated particles 
over the history of the PWN and in a region further offset from
the X-ray ``trail'' (\g359). If the centroid of the radio emission
represents the location where the pulsar was born, the separation
from the pulsar $\sim 4$ light years then gives an estimate of
the pulsar age of $\sim 1.3\times10^4 {\rm~years} (v_p/100 {\rm~km~s^{-1}}$,
where $v_p$ is the proper motion of the pulsar. In addition,
the high polarization (10-20$\%$), center-filled morphology, and relatively
flat non-thermal radio spectrum (Ho et al. 1985; $\S$4) are 
all consistent with the properties of a PWN.  
The ratio of the X-ray (0.2-10 keV) luminosity to radio 
luminosity ($L_x/L_r$$\sim$8) is also comparable to PWNe  
such as 3C58 (e.g., Helfand \& Becker 1987) and SNR G54.1+0.3 
(Lu et al. 2002). Thus, we conclude that the ram-pressure confined
PWN provides a unified explanation for the linear complex formed by
the X-ray source CXCGCS J174539.6-290413, the X-ray emission in \g359, and
the radio counterpart.

We have also tried to study whether there exists any X-ray spectral
evolution along \g359 with increase distance from the point source
CXCGCS J174539.6-290413. Because all the photons softer than 4 keV from
\g359 have been absorbed by ISM, we calculate the hardness ratios 
(the 5.5-9.0 keV counts divided by the 4.0-5.5 keV counts) of the two 
regions illustrated in Figure 1. The hardness ratio of the upper right 
region is 1.09$\pm$0.15 and that of the lower left region is
1.08$\pm$0.15, indicating that no spectral evolution along the length
has been detected. However, this is not evidence against a PWN scenario.
In a PWN, the X-ray spectrum
of the closest region (from the pulsar) is usually flatter than the
X-ray spectrum of the furthest region with a photon index difference
of $\sim$0.5 (e.g., Lu et al. 2002; Willingale et al. 2001). Such a
photon index difference corresponds to a hardness  difference of only 0.10,
if we assume that the X-ray spectrum of the furthest region is a power law
with a photon index of -2.0. Apparently, the quality of the current data is
not high enough to permit a spectral evolution study.

\subsection{G359.54+0.18}

There is no sign for any radio SNR emission near the X-ray filament \thread.
The NTF \thread~is clearly not associated with a SNR and exhibits the morphology
and physical properties of the NTFs. In addition, the X-ray emission 
is detected only in the middle of the northern of the two strands in \thread.
The southern filament is actually brighter in radio 
(Figure 5, see also Yusef-Zadeh et al. 1997), which is different from 
the strong correlation
between the surface brightnesses of the nonthermal X-ray 
and radio emission as seen in SNRs. 

Can G359.54+0.18 also be a PWN? There is no evidence for a point-like 
X-ray source that may represent a young energetic pulsar near \thread, 
and which traces the origin of the relativistic particles.
Of course, this lack of a point-like X-ray source
alone does not exclude the PWN possibility.
The source may be too faint to be detected with the existing data, or
the X-ray emission may be beamed away from us. The most serious 
difficulty of the PWN interpretation is the presence of the two
radio filaments and the fact that the X-ray emission is only 
associated with one of them. One possibility is that the pulsar is moving, 
and thus on its path, it illuminated the southern strand of \thread~in the recent past and is now 
located near the northern strand of the NTF. Therefore the lack of X-ray 
emission from the southern strand would then be due to the
synchrotron cooling of relativistic particles. The presence of 
into magnetic field flux tubes may be related to the synchrotron cooling 
instability of the PWN. But this mechanism does not explain why there 
are only two adjacent tubes. Furthermore, the radio spectral index of
this NTF, $\alpha$=-0.8$\pm$0.1 (Anantharamaiah et al. 1991), is 
substantially steeper than the typical for PWNe. 

Finally, let us examine the particle acceleration via magnetic reconnection.  
Magnetic reconnection is a source of inductive electric field, which 
can accelerate electric particles (Blandford 1994). Observational evidence
for particle acceleration by magnetic reconnection has been obtained by
investigating the hard X-ray emission of solar flares (e.g., 
Masuda et al. 1994). However, the actual acceleration process is 
complicated and has not been well understood so far (Blandford 1994).
Nevertheless, the co-existence of the two nearly parallel radio filaments 
makes the magnetic reconnection an attractive interpretation for \g359. 
The orientation of the magnetic field is 
found to be parallel to the filaments (Yusef-Zadeh et al. 1997), although
the direction of the field lines are unknown. If the lines run in 
opposite directions in the two filaments, the magnetic reconnection 
might have occurred in a close contact between the two filaments, 
leading to the acceleration of particles. This may provide a reasonable
explanation for the radio emission from the two filaments.
But the mechanism still does not explain why the X-ray emission coincides
with one of the filaments, not in-between them. Future, deeper X-ray 
observations are needed to make closer and better comparisons with the 
existing detailed radio data on this intriguing NTF. 

\acknowledgments
We thank S. Immler and R. Williams for their comments. This work is supported 
partially by NASA-grant SAO GO-12068X and NASA LTSA grant NAG5-7935.  
FJL also appreciates support of the Special Funds for Major State 
Basic Research Projects and the National Natural Science
Foundation of China.

\clearpage


\begin{figure}
\plotone{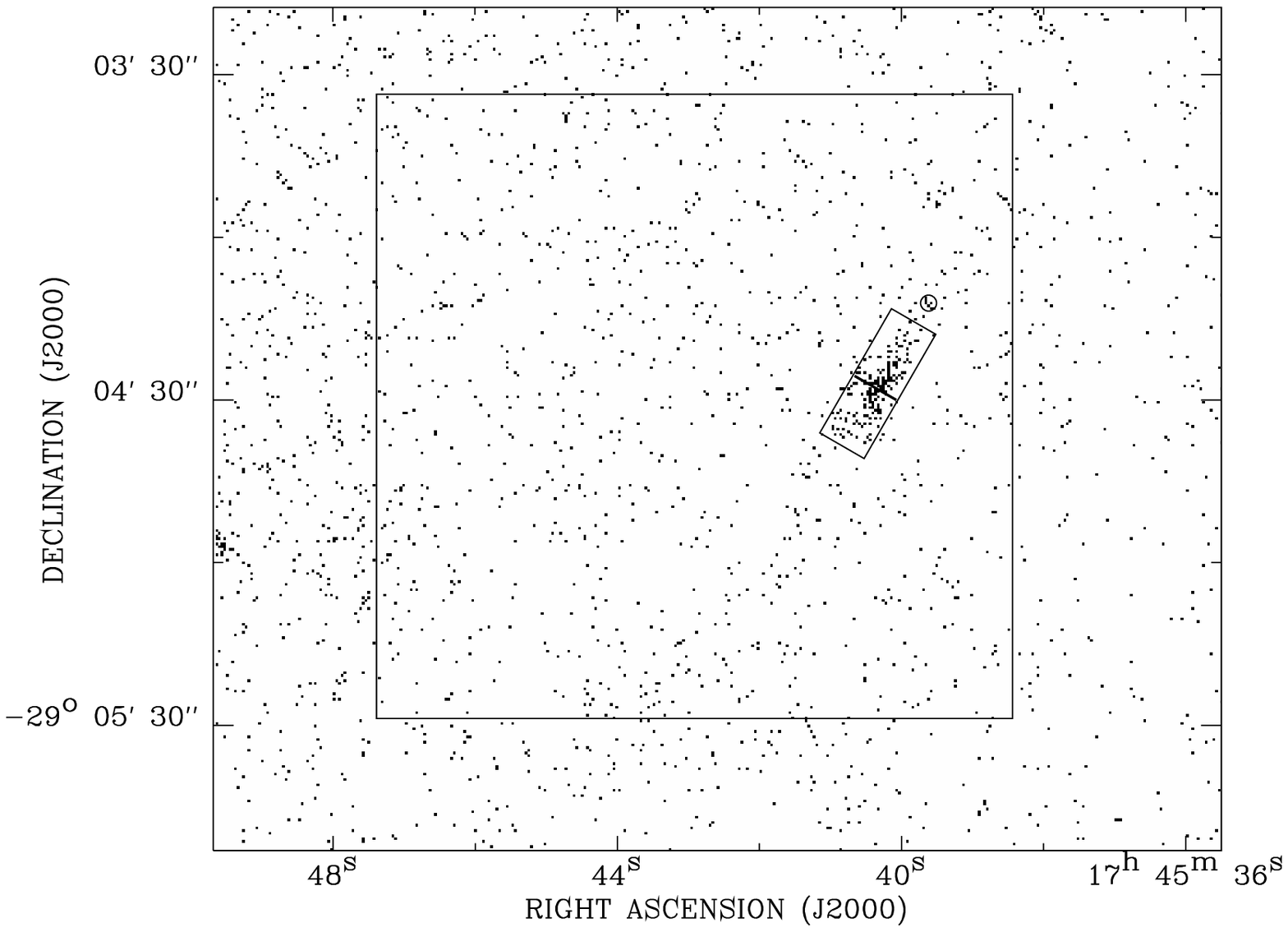}
\caption{4.0-9.0 keV {\sl Chandra} ACIS-I count distribution 
in the field of the X-ray wisp G359.89-0.08 (the small rectagular box).
The circle denotes a point-like source ( CXCGCS J174539.6-290413)
that may represent a pulsar, whereas the square big box 
outlines the region (with the small box and circle excluded) 
from which the background spectrum is extracted. The heavy solid line
devide G359.89-0.08 into two regions used to study the hardness ratio
variation.
\label{fig1}}
\end{figure}
\clearpage

\begin{figure}
\plotone{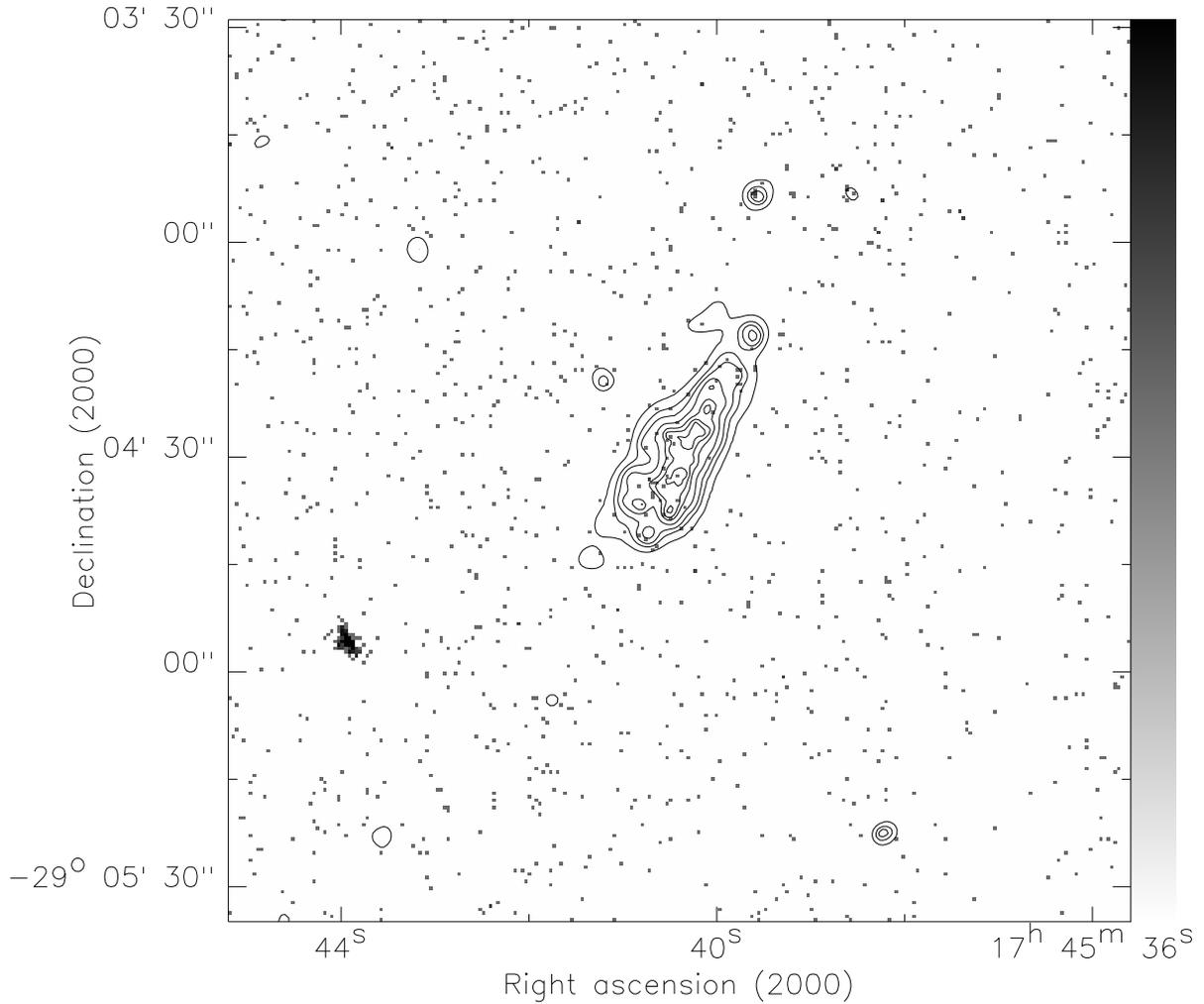}
\caption{The 0.5-4.0 keV {\sl Chandra} ACIS-I count map of the 
G359.889-0.082 region with the 4.0-9.0 keV intensity contours overlaid.
The intensity map represented by the contours has been smoothed with
an adaptive Gaussian filter to achieve counts-to-noise ratio of 4. The
contours are at levels of 0.33, 0.50, 0.66, 0.99, 1.65, 2.48, 
and 4.13 cts arcsec$^{-2}$.  Data from both OBS \#242 and OBS \#1561
are used.   
\label{fig2}}
\end{figure}

\begin{figure}
\plotone{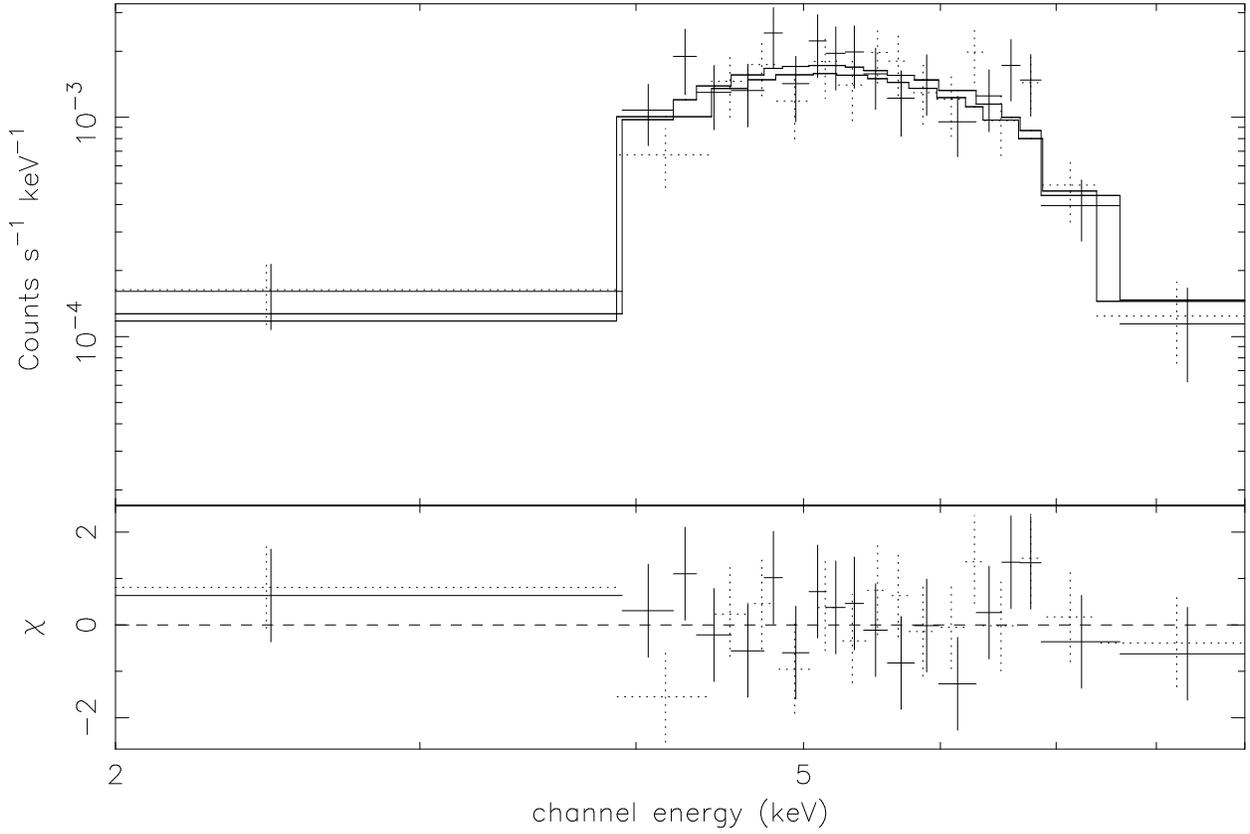}
\caption{The X-ray spectrum of \g359\ (shown in black) fitted a with a 
power-law model. The data represented by solid crosses are from OBS \#242,
and those represented by dotted crosses are from OBS \#1561.  
\label{fig3}}
\end{figure}

\begin{figure}
\plotone{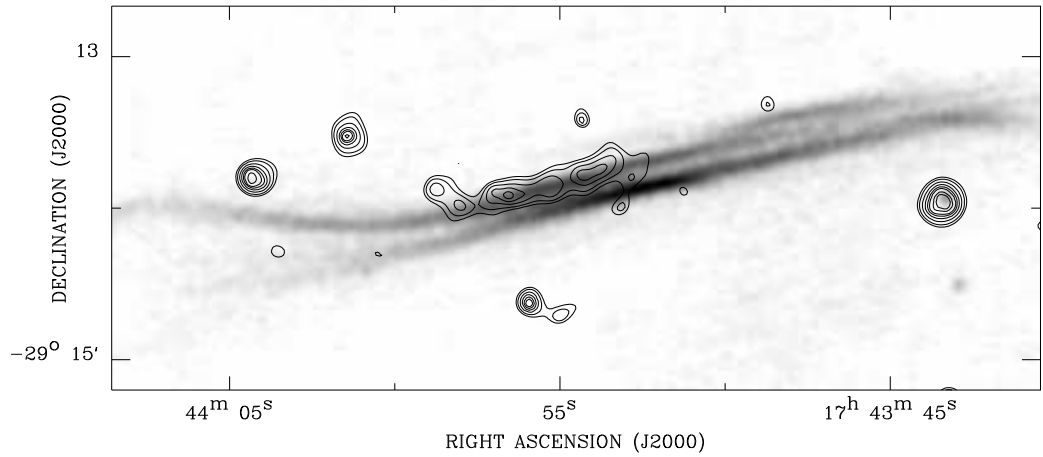}
\caption{A smoothed 2.0-9.0 keV {\sl Chandra} ACIS-I intensity map 
of the X-ray emission associated with G359.54+0.18, overlaid on the high resolution (4\arcsec) 
6 cm radio continuum image from Yusef-Zadeh et al. (1997).
\label{fig5}}
\end{figure}

\begin{figure}
\plotone{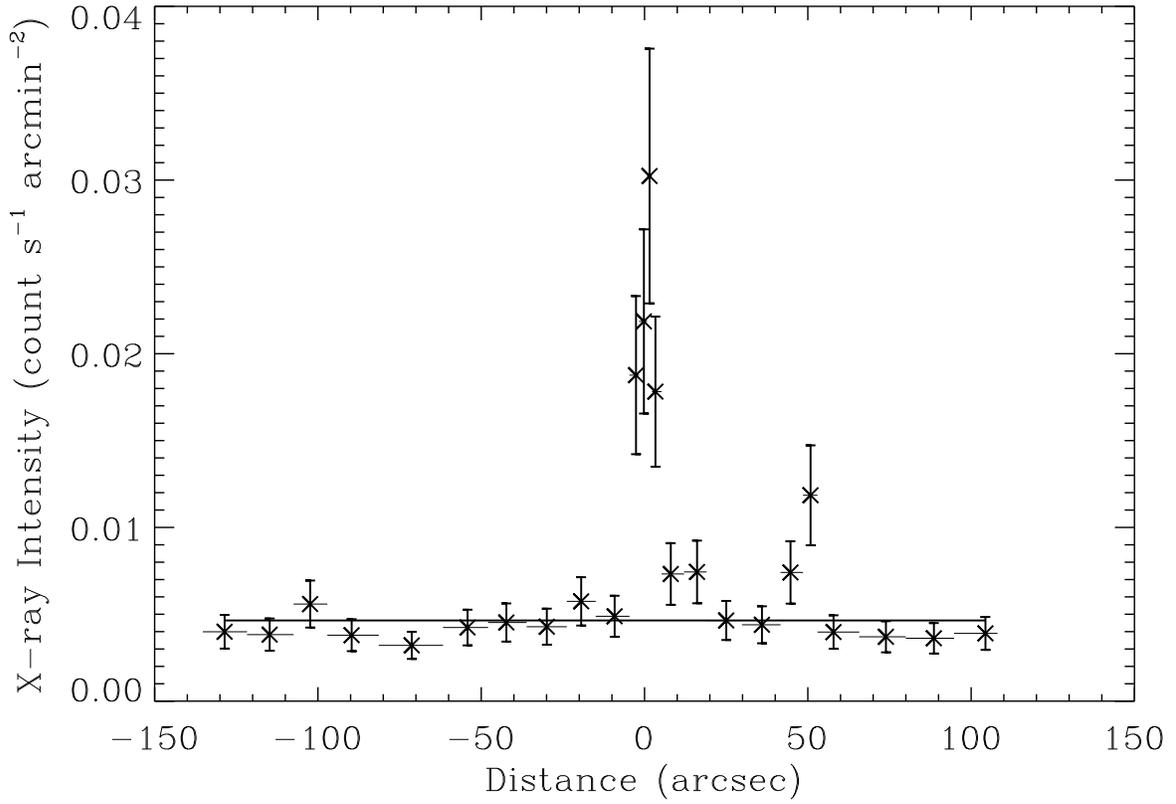}
\caption{A cross-cut of the X-ray Intensity from the raw counts 
image (counts s${-1}$ arcmin$^{-2}$) of the
G359.54+0.18 NTF. This plot shows that the X-ray emission in G359.54+0.18 is resolved to a 
width of $\sim$10\arcsec.
\label{fig5}}
\end{figure}

\begin{figure}
\plotone{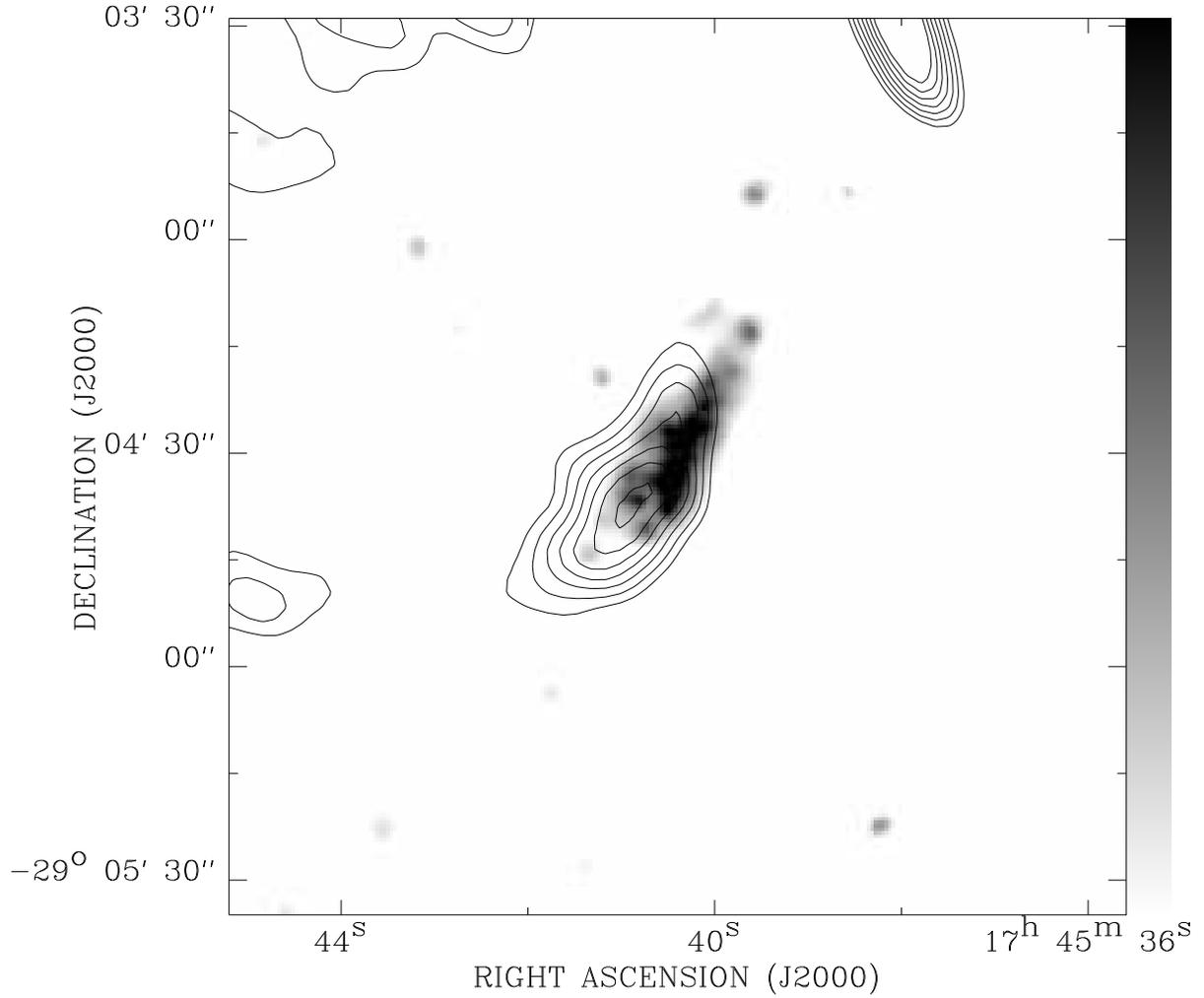}
\caption{The smoothed 4.0-9.0 keV {\sl Chandra} ACIS-I counts map
of \g359\ with the 20 cm radio contours overlaid.  The two sources in the 
very north of \g359 show much lower ISM absorption column densities 
compared to that to \g359.  The  color
bar changes logarithmically from 0.29 to 2.9 cts arcsec$^{-2}$. 
\label{fig6}}
\end{figure}

\begin{deluxetable}{lccccc}
\tabletypesize{\footnotesize}
\tablecaption{Model parameters of the X-ray spectral fitting of \g359 \label{tbl-1}}
\tablewidth{0pt}
\tablehead{
\colhead{Model}   & 
\colhead{$N_H$\tablenotemark{a}}   &
\colhead{Temperature(keV)/Photo Index}  & 
\colhead{$F_1$\tablenotemark{b}} &
\colhead{$F_2$\tablenotemark{b}} & 
\colhead{$\chi^2$/DOF\tablenotemark{c}}}
\startdata
 Thermal Plasma & 3.0$^{+0.5}_{-0.4}$ & 13.5($\geq$9.0)  & 0.3 & 1.4 & 20.0/34\\
 Power Law & 3.7$^{+1.4}_{-1.2} $ & 1.9$^{+1.3}_{-1.0}$ & 0.3 & 2.2 & 20.4/34\\
\enddata


\tablenotetext{a}{Column density in units of $10^{23}$cm$^{-2}$.}
\tablenotetext{b}{{$F_1$ and $F_2$ are absorbed and unabsorbed fluxes, respectively;
both in the 0.2-10.0 keV band and in units of 10$^{-12}$ ergs cm$^{-2}$ s$^{-1}$.}} 
\tablenotetext{c}{Degree of freedom.}
\tablecomments{The listed uncertainties are at the 90$\%$ confidence level.}

\end{deluxetable}
\clearpage

\end{document}